\documentclass[10pt,conference]{IEEEtran}
\IEEEoverridecommandlockouts
\usepackage{cite}
\usepackage{amsmath,amssymb,amsfonts}
\usepackage{algorithmic}
\usepackage{graphicx}
\usepackage{textcomp}
\usepackage{framed}
\usepackage{hyperref}
\usepackage{xcolor}
\def\BibTeX{{\rm B\kern-.05em{\sc i\kern-.025em b}\kern-.08em
    T\kern-.1667em\lower.7ex\hbox{E}\kern-.125emX}}

\newcommand{\rev}[1]{{\color{black}{#1}}}
\begin{document}

\title{Open Design Case Study - A Crowdsourcing Effort to Curate Software Design Case Studies}

\author{\IEEEauthorblockN{Chun Yong Chong}
\IEEEauthorblockA{\textit{School of Information Technology} \\
\textit{Monash University Malaysia}\\
Malaysia \\
chong.chunyong@monash.edu}
\and
\IEEEauthorblockN{Eunsuk Kang}
\IEEEauthorblockA{\textit{School of Computer Science} \\
\textit{Carnegie Mellon University}\\
USA \\
eunsukk@andrew.cmu.edu}
\and
\IEEEauthorblockN{Mary Shaw}
\IEEEauthorblockA{\textit{School of Computer Science} \\
\textit{Carnegie Mellon University}\\
USA \\
shaw@cs.cmu.edu}
}



\maketitle

\begin{abstract}

Case study-based learning has been successfully integrated into various courses, including software engineering education. In the context of software design courses, the use of case studies often entails sharing of real successful or failed software development. \rev{Using examples of real-world} case studies allows educators to reinforce the applicability and usefulness of fundamental design concepts, relate the importance of evaluating design trade-offs with respect to stakeholders' requirements, and highlight the importance of upfront design where students that lack industrial experience tend to overlook. However, the use of real-world case studies is not straightforward because 1.) there is a lack of open source repositories for real software design case studies and 2.) even if case studies are available, they are often reported without a standardized format, which may hinder the \rev{alignment between the case and the desired learning outcomes}. To address the lack of software design case studies for educational purposes, we propose the idea of Open Design Case Study, a repository \rev{to crowdsource, curate, and recruit other educators to contribute case studies for teaching software design courses}. The platform will also allow educators and students to share, brainstorm, and discuss design solutions based on case studies shared publicly on the repository. 

\end{abstract}


\section{Introduction}\label{sec:introduction}



``Software design'' refers to a set of activities that are involved in bridging the gap between requirements and an implementation, including domain modeling, design space exploration, architectural design and analysis, as well as module and code design. \rev{It may also involve discovering or refining requirements.} Design has significant impact on important qualities of the resulting system, such as modularity, robustness, maintainability, testability, and usability \cite{shaw2006golden}.

Despite its importance, software design can be a difficult topic to teach, partly due to the ``\rev{abstract} nature'' of high-level, abstract concepts in design compared to more concrete software artifacts (e.g., code), as reported by Galster and Angelov \cite{galster2016makes}. Unlike other software engineering activities such as software testing, where there are measurable targets (i.e. code coverage), developing a design solution (e.g., an architecture) requires a deep understanding of the often-messy problem domain and use cases, with multiple, alternative solutions that may not be easily comparable. Hence, it can be challenging for inexperienced students to grasp the concepts of software design, especially if it is taught without concrete examples to relate how such abstract concepts are relevant in a \rev{software engineering (SE)} lifecycle. 

The challenge of teaching software design is further aggravated by gaps between what is taught in universities and what is expected from the industry, as discussed by Garousi et al. \cite{garousi2019closing}. In their systematic review, the authors found that the curriculum in software design is of high importance and high gap, meaning that the software design curriculum requires the most attention with respect to the need for improvements in tertiary software engineering education programs. 

As such, there is an urgent need to improve the way we teach software design to undergraduate students. Multiple authors have proposed the use of case studies (fictitious or real-life examples) to help students to understand the importance of design choices when dealing with complex software systems with multiple stakeholder needs \cite{varma2005case, kurkovsky2015teaching, ouh2019applying}. \rev{Note that case studies mentioned in this paper refers to case-based pedagogical approach as discussed in the work by Delacey and Leonard \cite{delacey2002case}}. Although it has been shown to be useful, the effectiveness of the case study-based approach is highly dependent on the availability of good case studies. Unfortunately, the difficulty in obtaining sample design artifacts and documents to be used for teaching is one of the biggest obstacles for SE educators. \rev{Developers are often reluctant to share them because they are mostly proprietary, contain trade secrets, or bounded by non-disclosure agreements}; even when they are available, they are often scattered across institutions and course-specific repositories.

In this idea paper, we propose \textbf{\textit{``Open Design Case Study'' (ODCS)}} as a platform for SE educators to crowdsource, share, and propose new strategies to improve the software design syllabus. This initiative is inspired by the Open Case Studies \cite{breshock_2021},  a data science education platform where educators and practitioners share real-world examples and best practices in data science. We propose a template\footnote{\href{https://github.com/opendesigncasestudies/odcs-template}{https://github.com/opendesigncasestudies/odcs-template}} for contributors to share their real-world software design case studies \rev{on GitHub} to aid in curriculum design. Several ideas on how SE educators can leverage the ODCS repository are discussed in this paper.  

\section{Background}

\subsection{Challenges in Teaching Software Design}
One way to tame the complexity of a software system is to devise a high-level abstraction of the system that helps to decompose it into smaller, more manageable components and allows software engineers to easily identify the dependencies between the components \cite{shaw2005deciding, shaw2000software}. 

However, a major challenge in conveying the importance of software design to students who lack real-world experience is that the benefits of explicit design can be difficult to grasp due to the abstract nature of key design concepts such as modularity and information hiding. Thus, common questions arise, such as ``Why can't I start writing the program and decide the design later?'', ``Why do we need to follow these prescriptive steps when we are adopting agile methodologies?'', ``My program works fine with the given inputs, so why do I need to care about the design?''.  The answers to these questions can be especially hard to explain if the assignments are based on fictitious example with artificial requirements that are not grounded in real-world systems. 

\rev{Understanding how to compare design alternatives and evaluate trade-offs between them, usually is reflected through questions such as}: ``Should we use this design or pattern?'', ``Why microservice architecture instead of client-server?'', ``What makes this solution superior than the other?'', ``Is there a golden or standard solution for the choice of architecture?'' \cite{rupakheti2015teaching}.   
As such, it may not be sufficient to simply present a case study, but to deeply engage students in generating, comparing, and evaluating design alternatives and systematically arriving at a solution in consideration of the given requirements.


Shaw et al. \cite{shaw2005deciding} designed their Carnegie Mellon software design course around a set of core competencies (such as identifying the type and structure of the problem, and understanding business and economic constraints). The objectives of the course was to allow students to understand and make decisions based on both technical and contextual requirements from stakeholders and to select design solutions that genuinely focus on their needs. In order to achieve these core competencies, the use of real-world case studies is also noted as playing an important role. 


\subsection{Case Study-based Learning in Software Design Courses}
Recognizing the importance of real-world exposure and hands-on practices in software engineering education, Varma and Garg \cite{varma2005case} proposed the use of case studies-based learning in SE education. The goal of the approach is to present a multifaceted view of software engineering problems that closely resemble how it is practiced in real-life. However, the authors also acknowledged \rev{a significant} obstacle to such an approach, which is the limited availability of open and public SE case studies. They further highlight that most open case studies usually present hypothetical solutions and research cases that might not be realistic from a practitioners' point of view. 

Lieh and Irawan \cite{lieh2018teaching} shared a multi-year analysis of using case study-based learning to teach software design. They discussed the importance of imparting practical software design skills in a university setting and its relevance to the learner's environment. To evaluate the effectiveness of case study-based learning, the authors compared it with traditional problem-based learning, which does not have predefined goals and expected outcomes. Based on their findings, the authors found that case study-based learning is better suited for students with little to no industry experience, while problem-based learning is more preferred by working adults with experience in real-world software design problems. However, the authors highlight that the proposed approach requires substantial effort from educators to prepare the course materials, especially to curate relevant case studies.

\subsection{Awareness of the Need for an Open Repository of Software Design Case Studies}

A Bird of Feathers (BoF) session on teaching software design was held on May 26, 2022 part of ICSE in Pittsburgh, USA.\footnote{\href{https://conf.researchr.org/track/icse-2022/icse-2022-birds-of-a-feather}{https://conf.researchr.org/track/icse-2022/icse-2022-birds-of-a-feather}} The goal of the BoF was to share experiences and lessons learned from teaching design and to discuss new ideas for incorporating design into software engineering curriculums. There were approximately 40 to 50 attendees at the workshop.  The attendees were presented with the following questions for discussion:

\begin{itemize}
    \item How do we describe and present design principles to students?
    \item What in-class, hands-on activities can we provide to enhance learning?
    \item What are some examples of good and bad designs to study as case studies?
    \item How do we evaluate how well students learn design?
\end{itemize}

Some common design principles were mentioned by multiple participants as being taught in existing courses, including separation of concerns, locality, information hiding, GoF patterns, and architectural patterns. However, others brought up a number of principles or design dimensions that they believed were poorly taught or lacking in the existing curricula, particularly writing, reading, and asking critical questions about design documents.

One of the key challenges in teaching software design highlighted during the discussion was the difficulty in obtaining sample software designs and documents to be used for teaching, because developers are often reluctant to share them. To address this issue, multiple participants suggested the idea of developing a shared online repository to collect examples and case studies for teaching software design courses. 

\subsection{Preliminary findings}

Through a \rev{literature review of the related works}, we have found a lack of publicly available real-world software design case studies. Part of the reason is that most software design documents are proprietary and might potentially reveal trade secrets. Even if such case studies are available, they are usually reported independently as part of a particular research and are organized or reported differently. We believe that an obstacle to collecting and curating such design case studies is the lack of a generally-accepted template to put together all the necessary documents in a consistent form.

In the domain of Data Science education, having access to real-world case studies is also important for students to understand the need of applying different analytical models to interpret the data, based on different contexts. Therefore, the Johns Hopkins Data Science Lab team proposed Open Case Studies \cite{breshock_2021}, which serves as a platform for sharing and curating real-world case studies for the design of data science curriculum. The case studies are meant to be used in universities to help contextualize real-world data science problems. To encourage contributors and help facilitate the sharing of such case studies, the authors proposed a simple and extensible template to communicate the core information of each study. Guidelines are available to assist and streamline the publishing process. With the success of Open Case Studies, we identify an opportunity to adopt a similar approach in curating and crowdsourcing case studies for software design curriculum development. 

\section{Proposed Direction and Approach}\label{sec:direction}

In this idea Paper, we propose the Open Design Case Study (ODCS) project, an attempt to curate real-world software design case studies through crowdsourcing. The aim is to provide an open-source repository for SE educators to share and discuss software design case studies. 

To facilitate the sharing of open design case studies, we propose a template for structuring the case studies, as shown in Figure \ref{fig:template}. 

 \begin{figure}[t]
      {\vspace{-0.5em}}
      \begin{framed}
      \begin{flushleft}
      \scriptsize
      \noindent{\textbf{Open Design Case Study Template}}

        \begin{enumerate}
              \item Citation
              \item Title
              \item Objectives
              \item Stakeholder
              \item Requirements
                 \begin{enumerate}
                       \item Descriptions
                       \item Constraints
                       \item Quality Attributes
                 \end{enumerate}
             \item Environment
                \begin{enumerate}
                    \item Entities and Assumptions
                \end{enumerate}
             \item Design Solution(s)
             \item Outcome
                \begin{enumerate}
                    \item Success
                    \item Failure
                \end{enumerate}
            \item Lessons Learned
            \item Teaching Materials
                \begin{enumerate}
                    \item Suggested Usage
                \end{enumerate}
            \item Other Notes and Resources
 \end{enumerate}    
      \end{flushleft}
      \end{framed}
      {\vspace{-1em}}
      \setlength{\belowcaptionskip}{-0.5em}
      \caption{Proposed Open Design Case Study Template}
      \label{fig:template}
    \end{figure}

Note that the headings and sub-headings in Figure 1 are meant as a guideline, and we do not expect contributors to provide exactly the information mentioned in the template. This flexibility will also allow  organizations to selectively share only parts of a case study that they are able to.

\noindent{\textbf{\textit{Citation:}} If the case study is part of a published work or is hosted on other platform (i.e. Zenodo, Figshare, etc.), the contributors can choose to include the citation information related to the case study.}

\noindent{\textbf{\textit{Objectives:}} Contributors can include the project objectives in this section.}

\noindent{\textbf{\textit{Stakeholders:}} Contributors can provide information about the stakeholders involved in the case study (direct or indirect). Contributors can also choose to include annoymized information about the stakeholders if necessary.}

\noindent{\textbf{\textit{Requirements:}} To fully leverage the case studies for curriculum design, contributors are advised to include as much information as possible, including  functional requirements, constraints (technical, environment, or business), as well as the quality attributes that were significant in the project. If the Software Requirements Specification (SRS) document is available, contributors are advised to provide a link to access the document.}

\noindent{\textbf{\textit{Environment:}} Contributors can discuss the important entities and actors in the problem domain (e.g., users or physical components), including their properties, roles, and assumptions on how they interact with the system.}

\noindent{\textbf{\textit{Design Solution(s):}} Any design artifacts at different levels of granularity can be added here. For example, design best practices, design patterns, architectural patterns, UML diagrams (use cases, classes) etc. Contributors can choose to share the design artifacts if permissible; it would be ideal to have these. We acknowledged that some of the case studies might not have a proposed design solution. This creates an opportunity for crowdsourcing of design solutions, which will be discussed in the next section.}

\noindent{\textbf{\textit{Outcome:}} Contributors can discuss any available success or failure story based on the proposed design solution. It will be an added bonus to discuss the root cause of the success or failure, if applicable.}

\noindent{\textbf{\textit{Lessons Learned:}} Contributors can include retrospective section based on the case study. This can be useful for educators and students to discuss the lessons learned, such as approaches to adopt or avoid for a similar type of design problem.}

\noindent{\textbf{\textit{Teaching Materials:}} Contributors can provide a reusable teaching materials. They can also discuss the suggested usage of the case study, including 1.) use to teach a design concept, 2.) use as assignment specification, 3.) use as a case study to avoid certain bad practices.}


To put the ODCS template in practice and demonstrate its applicability, we have curated several case studies that are publicly available on GitHub.\footnote{\href{https://github.com/opendesigncasestudies}{https://github.com/opendesigncasestudies}} In particular, these projects are real-world case studies discussed in the work by Sommerville \cite{sommerville2015} and the London Ambulance Service case study~\cite{finkelstein96}. 

Contributors are encouraged to utilize our published ODCS template\footnote{\href{https://github.com/opendesigncasestudies/odcs-template}{https://github.com/opendesigncasestudies/odcs-template}} to prepare additional the case studies. Once completed, \rev{they can add their case studies to the centralized repository to publish them}. 

\subsection{Suggested Usage of Open Design Case Study by Educators}


\subsubsection{Case studies as a medium to introduce basic design concepts}
The benefits of using case studies to teach and impart basic design concepts has been reported in multiple studies \cite{kurkovsky2015teaching,tao2006work}. Bolinger et al.\cite{bolinger2011connecting}, described how case studies can reinforce abstract concepts (design patterns, architecture, etc.), demonstrate the nature of real client interactions (conflicting requirements, prioritization of requirements, etc.), and showcase the relevance of soft skills to students that lack significant practical experience.

While design patterns are usually used in concert to solve complex software engineering problems in practice, they are usually taught using independent examples in universities \cite{tao2006work}. By leveraging real-world case studies, educators can discuss and relate the dynamics of design decision making when dealing with real-world software systems. Instead of using dummy examples, educators can provide real examples on how design patterns can be used to solve real world problems. \rev{However, we do acknowledge that the usage of design case studies may more easily work with technical decisions rather than holistic design decisions.}

\subsubsection{Use of case studies for curriculum development}
Regurgitating a detailed real-world case study as it is might be boring and hard to absorb by the students who do not appreciate the perspective of the original software developers. When the contents are customized and localized to fit the students' level of expertise, it will be easier for the students to understand the concepts in a way that is consistent with the fundamental design knowledge that is being taught to them \cite{bolinger2011connecting}. \rev{We encourage instructors who customize their content (design case studies) to share their results with the wider SE educator community on the ODCS GitHub page.}

As such, the usage of case studies can be used to drive curriculum development by allowing educators to use the case studies to design teaching materials that are more industry relevant \cite{bolinger2011connecting}. In-class activities such as tutorials or labs can be redesigned to revolve around the selected case studies to reinforce the understanding of abstract concepts. Design walkthroughs can also help students to understand the rationale behind the design decisions. Such interactive activities are opportunities and innovations that are aligned with the concept of flipped classroom, which has been proven to improve students' learning experience \cite{gren2020flipped}. 

\subsubsection{Platform to discuss and crowdsource design solutions} \label{sec:crowd}

As discussed in Section \ref{sec:direction}, we understand that not all case studies come with a proposed design solution due to different factors (trade secrets, incomplete case studies, etc.). Even when design solutions are available, there is always room to propose alternative design that can help address the functional and non-functional requirements from a different perspective.

As such, the proposed ODCS can be used as a platform to crowdsource design solutions among SE educators and students. In a more controlled teaching and learning environment, educators can fork a particular case study and instruct their students to submit pull requests to submit their design solutions. The design solutions can be further improved and refined over time. 

Apart from that, educators can create use cases that are more customized to the students' learning environment by scaffolding from the real case studies published on ODCS. This strategy can be used for conducting long-running projects where the requirements can evolve over multiple semesters. 

\section{Plans to Evaluate Open Design Case Study}

As an initial validation, we have put the proposed ODCS template (Figure \ref{fig:template}) to actual use by converting some of the open real-world case studies into the proposed format, as discussed in Section \ref{sec:direction}. In the process, it has helped us refine the overall structure and subsections of the proposed template to be more well suited for SE education, focusing on the ``Why'' and ``How'' factors in design decision making. For instance, the ``Outcome'' and ``Lesson Learned'' sections in the proposed template help provide opportunities for educators to run retrospective and discussion sessions with the students, and to teach them how to critique the proposed design solution(s). We plan to further enrich the ODCS by incorporating more open case studies on the GitHub repository. 

\begin{figure}[ht!]
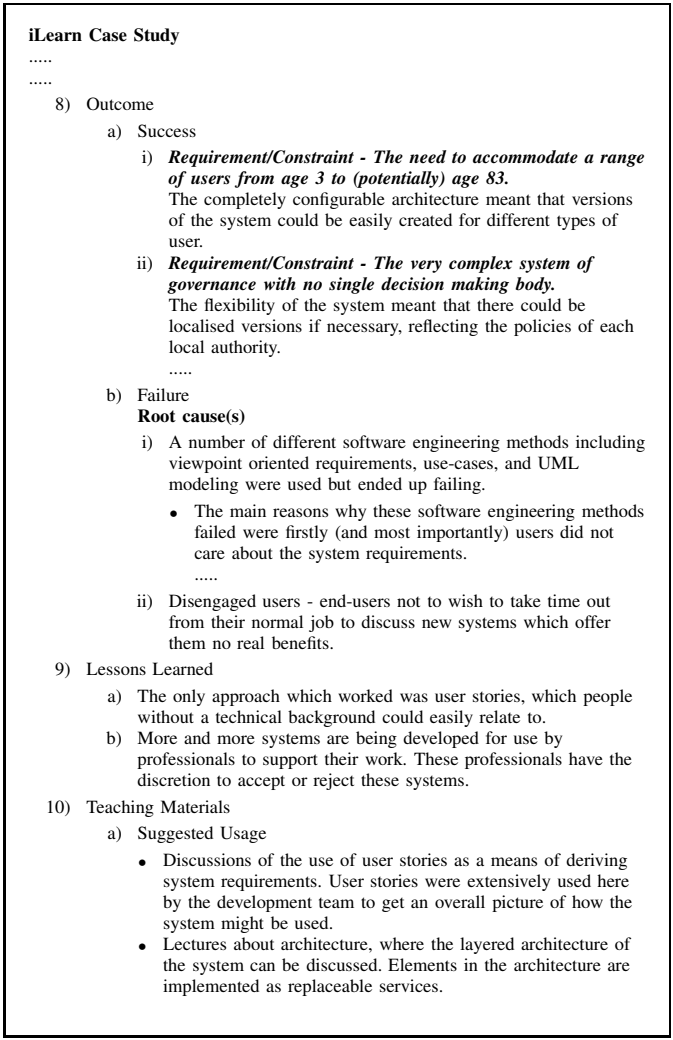

      {\vspace{-0.5em}}
      \begin{framed}
      \begin{flushleft}
      \scriptsize
      \noindent{\textbf{iLearn Case Study}}
        \newline
        .....
        \newline
        .....
        \begin{enumerate}
         \setcounter{enumi}{7}
             \item Outcome
                \begin{enumerate}
                    \item Success
                    \begin{enumerate}
                        \item \textit{\textbf{Requirement/Constraint - The need to accommodate a range of users from age 3 to (potentially) age 83.}}
                        \newline
                        The completely configurable architecture meant that versions of the system could be easily created for different types of user.
                        \item \textit{\textbf{Requirement/Constraint - The very complex system of governance with no single decision making body.}}
                        \newline
                        The flexibility of the system meant that there could be localised versions if necessary, reflecting the policies of each local authority.
                          \newline
                          .....
                    \end{enumerate}
                    \item Failure
                    \newline
                    \textbf{Root cause(s)}
                    \begin{enumerate}
                        \item A number of different software engineering methods including viewpoint oriented requirements, use-cases, and UML modeling were used but ended up failing.
                        \begin{itemize}
                            \item The main reasons why these software engineering methods failed were firstly (and most importantly) users did not care about the system requirements.
                            \newline
                            .....
                        \end{itemize}
                        \item Disengaged users - end-users not to wish to take time out from their normal job to discuss new systems which offer them no real benefits.
                    \end{enumerate}
                \end{enumerate}
            \item Lessons Learned
            \begin{enumerate}
                \item The only approach which worked was user stories, which people without a technical background could easily relate to.
                \item More and more systems are being developed for use by professionals to support their work. These professionals have the discretion to accept or reject these systems.

            \end{enumerate}
            \item Teaching Materials
                \begin{enumerate}
                    \item Suggested Usage
                    \begin{itemize}
                        \item Discussions of the use of user stories as a means of deriving system requirements. User stories were extensively used here by the development team to get an overall picture of how the system might be used.
                        \item Lectures about architecture, where the layered architecture of the system can be discussed. Elements in the architecture are implemented as replaceable services.
                    \end{itemize}
                \end{enumerate}
            
 \end{enumerate}    
      \end{flushleft}
      \end{framed}
      {\vspace{-1em}}
      \setlength{\belowcaptionskip}{-0.5em}
      \caption{Snippet from the iLearn Case Study}
      \label{fig:ilearn}
    \end{figure}

An example of the case study reported using the proposed template is shown in Figure \ref{fig:ilearn}, where the figure shows a snippet of the iLearn Case Study\footnote{\href{https://github.com/opendesigncasestudies/iLearn-IanSommerville}{https://github.com/opendesigncasestudies/iLearn-IanSommerville}} from Sommerville \cite{sommerville2015}. Educators can use the ``Success'' or ``Failure'' stories to highlight design decisions to adopt or avoid based on the outcome of real-world systems. The ``Teaching Materials'' section provide some suggested usage of the case studies to facilitate the sharing of these real-world examples.  

We also plan to conduct qualitative studies to collect feedback from SE educators about the usage of ODCS, its benefits in their own classrooms as well as limitations, and use the collected data to further improve the template. Some of the plans to conduct qualitative studies include focus group discussions, a survey and questionnaire from conference attendees (at software engineering venues), as well as an opt-in survey for SE educators who plan to use or contribute towards the ODCS. We also plan to evaluate whether the granularity of the proposed template is too content heavy, or too abstract or vague to be useful for SE educators. \rev{We invite readers who are interested to participate in the qualitative studies to contact the authors of this paper.}


In addition, for educators who plan to use OCDS for teaching software design courses in universities, they can consider conducting a comparative study to evaluate the effectiveness of case study-based learning against traditional learning methods, similar to the work by Lieh and Irawan \cite{lieh2018teaching}. The findings from the comparative study can help pave way to improve the pedagogical approach for case study-based learning in software design courses. 

\section{Related Work}

In this section, we discuss some existing attempts at collecting SE case studies or datasets.

There have been multiple attempts at curating open source repositories and datasets to help improve the replicability and reliability of software engineering research. One of the earliest attempt is the PROMISE (PRedictOr Models In Software Engineering) repository \cite{Sayyad-Shirabad+Menzies:2005} which has been widely used in different domain of software engineering including defect prediction, software cost estimation, and reuse prediction. 

The Mining Software Repositories (MSR) research community is one of the strongest advocates of publicly available datasets where projects such as GitTorrent \cite{gousios2012ghtorrent}, FLOSSMole \cite{howison2006flossmole}, and FLOSSMetrics \cite{herraiz2009flossmetrics} are widely used in conducting research related to open source software development. While open source software development platforms such as GitHub are frequently used by the MSR communities to conduct research, Kalliamvakou et al. \cite{kalliamvakou2016depth,kalliamvakou2014promises} found that without proper scrutinization and guidelines, research that relies on such open source repositories might not be reliable because the majority of the projects hosted on GitHub are personal and inactive. 

On the other hand, open datasets for past software defects are crucial to enable researchers to evaluate their defect prediction models. Datasets such as Defects4J \cite{just2014} and Bug prediction dataset \cite{DAmb2010a} are some of the most popular open source datasets for researchers in the domain of bugs/defects prediction. These kinds of open source defect datasets helps researchers working in the relevant domain to benchmark their proposed models against the state-of-art approaches in a objective and fair manner. 

In this section, we discuss some attempts to collect SE case studies, datasets, or educational materials. Software engineering educators are usually willing to share their educational materials. In most cases this sharing centers on materials for a specific course.  A project to support collaborative development of material in a new topic area was initiated for an emerging area of ``economics-driven software engineering'' in 2002 (a period when the dominate open source repository was Sourceforge). Shaw et al. \cite{shaw2003courseforges} reported on the CourseForges project, which provided an interactive platform for faculty to share the effort of developing course materials under ground rules modeled on open source principles with Creative Commons copyright. Although a body of material was shared, there was little collaborative evolution and the project faded. The  main reasons for the failure seem to be the lack of a clear structure for the level, granularity, and organization of the material and some concerns about whether faculty effort on shared educational artifacts would be appropriately recognized.   The former issue is addressed here by the ODCS template; as to the latter, we believe that educational contributions have a higher profile now than they did two decades ago.


\section{Conclusion and Future Work}

In this idea paper, we propose the Open Design Case Study (ODCS) to provide a platform for SE educators to curate and obtain real-world software design case studies. The goal of the initiative is to build a community based on crowdsourcing efforts that help mold the future of software design curriculum. By proposing the ODCS template, we aim to use it as a guideline to provide a focused view of case studies that reveal how design decisions were made, how design solutions were proposed to fulfill different quality requirements, and what are the success or failure stories that students can learn from it. 

We do acknowledge that there are some risks of using the proposed template due to incomplete software design artifact/documentations, or lack of design solution. To help mitigate the risk, we have discussed in Section \ref{sec:crowd} on how students or educators can utilize the ODCS platform as a way to propose and refine incomplete design artifacts or solutions. 

Finally, as part of the future work to improve the readability of the design artifacts and delivery of contents to a wider community, we plan to use Jekyll\footnote{\href{https://jekyllrb.com/}{https://jekyllrb.com/}} to convert the current \texttt{MARKDOWN} contents to static websites. 

\bibliographystyle{IEEEtran}
\bibliography{references}

\end{document}